\documentclass[aps, prd, twocolumn, 
showpacs, showkeys, 
print, preprintnumbers, nofootinbib,
superscriptaddress 
]{revtex4-2}
\usepackage[utf8]{inputenc}
\usepackage[T1]{fontenc}
\usepackage{lmodern}
\usepackage{microtype}
\usepackage[english]{babel}
\usepackage{amssymb,amsbsy,amsmath,amsfonts,amsthm}

\usepackage{yhmath} 
\usepackage{slashed}
\usepackage{bm}
\usepackage{times}
\usepackage[dvipsnames,table,xcdraw]{xcolor}
\definecolor{lightyellow}{RGB}{255,250,205}
\usepackage{graphicx}
\usepackage{epstopdf}
\usepackage{tabularx}
\usepackage{multirow}
\usepackage{longtable}
\usepackage{hyperref}
\hypersetup{colorlinks=true, linkcolor=blue, 
citecolor=cyan, urlcolor=red }
\usepackage{url}

\usepackage{orcidlink}
\allowdisplaybreaks

\begin{document}

\title{Phenomenological model for the $\gamma^*\gamma \to \eta \pi^+ \pi^-$ reaction in the $f_1(1285)$ energy region}
\author{Xiu-Lei Ren\orcidlink{0000-0002-5138-7415}}
\affiliation{Helmholtz Institut Mainz,   D-55099 Mainz, Germany }

\author{Igor Danilkin\orcidlink{0000-0001-8950-0770}}
\affiliation{Institut f\"ur Kernphysik \& PRISMA$^+$  Cluster of Excellence, Johannes Gutenberg Universit\"at,  D-55099 Mainz, Germany}

\author{Marc Vanderhaeghen\orcidlink{0000-0003-2363-5124}}
\affiliation{Institut f\"ur Kernphysik \& PRISMA$^+$  Cluster of Excellence, Johannes Gutenberg Universit\"at,  D-55099 Mainz, Germany}

\begin{abstract}
Motivated by the ongoing analysis by the BESIII Collaboration on the single-tagged $e^+e^-\to e^+e^- \eta \pi^+\pi^-$ reaction, we present a phenomenological study of the diphoton fusion to $\eta \pi^+ \pi^-$, focusing on the production mechanism of the $f_1(1285)$ resonance. Contributions from the $f_1(1285)\to a_0(980)^\pm\, \pi^\mp$ and $f_1(1285)\to \sigma/f_0(500)\, \eta$ channels are included without introducing free parameters within an effective Lagrangian approach. Assuming the destructive interference between the amplitudes, we predict the invariant mass distributions, angular distributions, and total cross sections of the $\gamma^*\gamma \to \eta\pi^+\pi^-$ process, which will be tested by the forthcoming BESIII measurements. 

\end{abstract}
\maketitle
\date{\today}


\section{Introduction}
Photon-photon fusion is considered to be a clean way to explore the inner structure of mesons with positive C-parity. The importance of this process is heightened by its relevance to the physics of the Muon $g-2$.  
In the first edition of the white paper by the Muon $g-2$ Theory Initiative~\cite{Aoyama:2020ynm,Aoyama:2012wk,Aoyama:2019ryr,Czarnecki:2002nt,Gnendiger:2013pva,Davier:2017zfy,Keshavarzi:2018mgv,Colangelo:2018mtw,Hoferichter:2019mqg,Davier:2019can,Keshavarzi:2019abf,Kurz:2014wya,Melnikov:2003xd,Masjuan:2017tvw,Colangelo:2017fiz,Hoferichter:2018kwz,Gerardin:2019vio,Bijnens:2019ghy,Colangelo:2019uex,Blum:2019ugy,Colangelo:2014qya}, the experimental measurements of the photon-photon fusion to three mesons, such as $\gamma^{(*)}\gamma^* \to 3\pi$, $K\bar{K}\pi$, $\eta\pi\pi$, are listed as the priority in order to provide the necessary theoretical input for the data-driven approach: the transition form factors (TFFs) of the axial-vector mesons $a_1(1260)$, $f_1(1285)$, and $f_1(1420)$), as well as tensor mesons (e.g. $a_2(1232)$). 

The $f_1(1285)$ TFFs have been studied in previous phenomenological studies~\cite{Schuler:1997yw,Melnikov:2003xd,Jegerlehner:2017gek}. Two recent parameterizations of $f_1(1285)$ TFFs have been proposed~\cite{Roig:2019reh,Leutgeb:2019gbz} based on the resonance chiral theory and the holographic model. The asymptotic behavior of axial-vector TFFs has been derived in Ref.~\cite{Hoferichter:2020lap}, which is incorporated in a vector-meson dominance inspired parameterization of $f_1(1285)$ TFFs~\cite{Zanke:2021wiq,Hoferichter:2023tgp} . However, due to the lack of the direct experimental data, the $f_1(1285)$ TFFs are not well determined.  

Currently, the BESIII Collaboration has performed feasibility studies on the $\gamma\gamma^* \to \eta\pi^+\pi^-$~\cite{Becker_MasterThesis}, $K^+K^-\pi^0$ processes~\cite{Effenberger_MasterThesis}, and the analysis  of $\gamma^{(*)}\gamma^* \to \pi^+\pi^-\pi^0$ is planned. To align with those experimental progresses, we have conducted phenomenological studies of the $\gamma\gamma \to \pi^+\pi^-\pi^0$~\cite{Ren:2022fhp} and the $\gamma\gamma^*\to K^\pm \bar{K}^{*\mp}(892)\to K^+K^-\pi^0$~\cite{Ren:2024nuf} reactions within an effective Lagrangian approach. The proposed models could serve as input for the Monte Carlo (MC) generators for the ongoing data analysis of BESIII.  
In this paper we present a theoretical study of the $\gamma^*\gamma \to  \eta\pi^+\pi^-$ process, which is the prime reaction to extract the TFFs of the lowest-lying axial-vector meson $f_1(1285)$, given the $f_1(1285)\to \eta \pi\pi$ branching ratio is $\sim 52.2\%$~\cite{ParticleDataGroup:2024cfk}. Besides the $f_1(1285)$ production mechanism, there is also a very close pseudoscalar $\eta(1295)$, which could theoretically be produced in the diphoton fusion to $\eta\pi\pi$ process. However, this state was not observed in previous experiments of the untagged measurement of $\gamma\gamma \to  \eta\pi^+\pi^-$. Apart from the already relatively well studied $\eta^\prime(958)$ state, also the $\eta(1405)$ state can be produced in $\gamma^*\gamma\to\eta\pi^+\pi^-$ reaction. Given the large mass differences, these narrow resonances $\eta'(958)$, $f_1(1285)$, and $\eta(1405)$ are expected to be well separated in the $\eta\pi^+\pi^-$ mass spectrum. Therefore, the $\gamma^*\gamma \to  \eta\pi^+\pi^-$ presents an ideal process to investigate the $f_1(1285)$ state. 

In the 1980s, Mark II~\cite{Gidal:1987bn} and TPC/Two Gamma~\cite{TPCTwoGamma:1988izb} Collaborations measured the $e^+e^-\to e^+e^-\eta\pi^+\pi^-$ reaction with tagged and untagged final-state electrons, and observed the $f_1(1285)$ state in the tagged two-photon fusion process. Subsequently, the L3 Collaboration~\cite{L3:2000gjc,L3:2001cyf} also reported the production of the $f_1(1285)$ state 
in the same process, and measured the branching fraction $\Gamma(f_1(1285)\to a_0(980)\pi)/\Gamma(f_1(1285)\to \eta\pi\pi)$. However, L3 Collaboration used an indirect method, which relied on the comparison with the Monte Carlo distributions, to extract the momentum transfer. Until now, no new experimental research on this process has been reported. 
Recently, a feasibility study of the single-tagged $e^+e^-\to e^+e^-\eta\pi^+\pi^-$ reaction in the $f_1(1285)$ energy region was performed based on BESIII data~\cite{Becker_MasterThesis}. In this analysis, the ``GGResRC'' MC generator~\cite{Druzhinin:2014sba} was used to interpret the experimental data. As in this generator any interference among the helicity amplitudes of different channels is neglected, a precision extraction of the $f_1(1285)$ TFFs is hampered in such approach.  

\begin{figure}[t!]
  \includegraphics[width=0.45\textwidth]{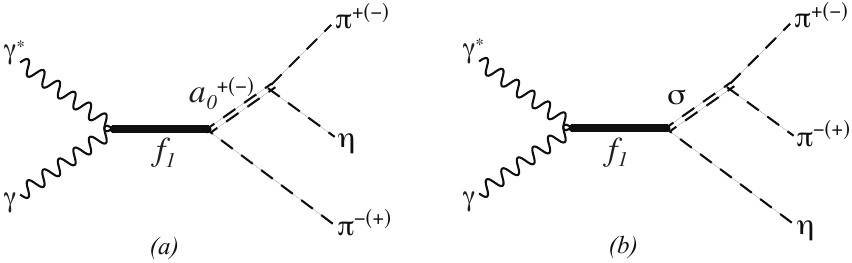}
  \caption{Feynman diagrams of the $\gamma^*\gamma\to \eta\pi^+\pi^-$ process via the $f_1(1285)\to a_0^{\pm}(980) \pi^\mp$ (a) and $f_1(1285)\to \sigma/f_0(500) \eta$ (b) decays.}
  \label{Fig:Feyn}
\end{figure}

To address this issue, we propose in this work a phenomenological model focusing on the $f_1(1285)$ production in the $s$-channel. As illustrated in Fig.~\ref{Fig:Feyn}, the two major decay processes $f_1(1285)\to a_0(980)^\pm\pi^\mp$ and $f_1(1285)\to \sigma/f_0(500) \eta$ are considered. Within the effective Lagrangian approach, we determine the couplings via the relevant decay widths. Through the destructive interference between the amplitudes of both channels, as hinted by the L3 data~\cite{L3:2001cyf}, we present our predictions for the invariant mass distributions, angular dependence, and total cross sections of the $\gamma^*\gamma \to \eta\pi^+\pi^-$ reaction. 
Additional potential mechanisms, such as the $\rho^0$-exchange in the $t$- and $u$-channels of the $\gamma^* \gamma \to \sigma \eta \to\eta\pi^+\pi^-$ process, are expected to be suppressed due to the heavy vector meson exchange. 
Moreover, the decay mechanism for $f_1(1285)$ into $\eta\pi\pi$ is also possible via a triangle $K^*\bar{K}K$ one-loop rescattering~\cite{Debastiani:2016xgg,Du:2021zdg}. The latter  contribution is however subdominant in the energy region of the $f_1(1285)$ excitation compared with the large tree diagram contribution~$f_1(1285)\to a_0\pi\to \eta\pi\pi$ as shown in Ref.~\cite{Debastiani:2016xgg}. 

The paper is organized as follows: In Sec.~\ref{SecII}, we present the amplitude of $\gamma^*\gamma \to \eta\pi^+\pi^-$ in our phenomenological model within the effective Lagrangian approach. In Sec.~\ref{SecIII}, the predictions of the polarized (differential) cross sections and angular distributions of $\gamma^*\gamma \to  \eta\pi^+\pi^-$ are shown in the low $Q^2$ region of BESIII measurement and the sensitivity to the $f_1(1285)$ TFFs is studied. Finally, we summarize the main results in Sec.~\ref{SecIV}.

\section{Framework}\label{SecII}
In this section, we formulate the Feynman amplitudes of the $\gamma^*(q_1,\lambda_1)+\gamma(q_2,\lambda_2)\to \eta(p_{\eta}) + \pi^+(p_{\pi^+}) + \pi^-(p_{\pi^-})$ reaction in our minimal model using the effective Lagrangian approach. The kinematical invariants of this $2\to 3$ process are defined by  
\begin{equation}
\label{Eq:stu}
\begin{aligned}
	&s =W^2 = (q_1+q_2)^2 = 2q_1\cdot q_2 - Q_1^2,\\
	&t = (q_1 - p_{\eta}-p_{\pi^+})^2,\\
	&u = (q_2 - p_{\eta}-p_{\pi^+})^2,\\
 	&M_{\pi^\pm\eta}^2 = (p_{\pi^\pm}+p_{\eta})^2,\\
	&M_{\pi^+\pi^-}^2 = (p_{\pi^+}+p_{\pi^-})^2\,
\end{aligned}
\end{equation}
with $q_1^2=-Q_1^2$ and $q_2^2=0$\,.

\subsection{Transition amplitude of $\gamma^* \gamma\to f_1(1285)$}

The transition amplitude of the $\gamma^*(q_1,\lambda_1)+\gamma^*(q_2,\lambda_2)\to A(P_A,\Lambda_A)$ reaction, where $A$ represents the axial-vector meson with momentum $P_A=q_1+q_2$ and helicity $\Lambda_A$, is parametrized by the three-independent structures~\cite{Poppe:1986dq,Schuler:1997yw},
\begin{equation}\label{Eq:Amp_gagaf1}
\begin{aligned}
\mathcal{M}_{A\gamma^*\gamma^*} 
& =i e^2\varepsilon_\mu(q_1,\lambda_1)\varepsilon_\nu(q_2,\lambda_2)\varepsilon^{\omega*}(P_A,\Lambda_A) \,  \epsilon_{\rho\sigma\tau\omega} \\
&\times\biggl[R^{\mu\rho}(q_1,q_2) R^{\nu\sigma}(q_1,q_2) (q_1-q_2)^\tau \\
&\qquad \times \frac{\nu}{M_{A}^2} F^{TT}_{A\gamma^*\gamma^*}(Q_1^2,Q_2^2) \\
	&\quad + R^{\nu\rho}(q_1,q_2)\biggl( q_1^\mu + \frac{Q_1^2}{\nu}q_2^\mu\biggr) q_1^\sigma q_2^\tau \\
 &\qquad \times \frac{1}{M_{A}^2} F^{LT}_{A\gamma^*\gamma^*}(Q_1^2,Q_2^2) \\
	&\quad + R^{\mu\rho}(q_1,q_2) \biggl(q_2^\nu + \frac{Q_2^2}{\nu}q_1^\nu\biggr) q_2^\sigma q_1^\tau \\
 &\qquad \times \frac{1}{M_{A}^2} F^{TL}_{A\gamma^*\gamma^*}(Q_1^2, Q_2^2) \biggr],
\end{aligned}	
\end{equation}
where the three TFFs  $F_{f_1\gamma^*\gamma^*}^{(TT,LT,TL)}(Q_1^2,Q_2^2)$, which are functions of both photons virtualities, describe the inner structure of the $f_1(1285)$ resonance. The superscript $TT$ denotes the fusion of two transverse photons, while $TL$ ($LT$) denotes the first photon in the fusion process being transverse (longitudinal) while the second photon being longitudinal (transverse), respectively. In Eq.~\eqref{Eq:Amp_gagaf1} $\varepsilon_{\mu}(q_i,\lambda_i)$ with $i=1,2$ are the polarization vectors of the incoming photons with helicity $\lambda_i$. 
The symmetric transverse tensor is defined as 
\begin{equation}
\begin{aligned}
  R^{\mu\nu}(q_1,q_2) &\equiv -g^{\mu\nu} + \frac{1}{X} \Biggl[ \nu(q_1^\mu q_2^\nu + q_2^\mu q_1^\nu) \\
  &\qquad\qquad   + Q_1^2 q_2^\mu q_2^\nu + Q_2^2 q_1^\mu q_1^\nu \Biggr],
\end{aligned}
\end{equation}
with the virtual photon flux factor 
\begin{align}
&X\equiv(q_1\cdot q_2)^2 - q_1^2 q_2^2 = \nu^2 -Q_1^2Q_2^2,\\
&\nu \equiv q_1\cdot q_2= \frac{W^2+Q_1^2+Q_2^2}{2}\,.
\end{align}
In the present work, we focus on the single virtual case, i.e. $\gamma^*(q_1,\lambda_1)\gamma(q_2,\lambda_2) \to f_1(1285)$. For this process, the above amplitude reduces as: 
\begin{equation}\label{Eq:Amp_gagaf1_Q12}
\begin{aligned}
& \mathcal{M}_{f_1\gamma^*\gamma} 
=i e^2\varepsilon_\mu(q_1,\lambda_1)\varepsilon_\nu(q_2,\lambda_2)\varepsilon^{\omega*}(q_1+q_2,\Lambda_{f_1}) \, \epsilon_{\rho\sigma\tau\omega}\\
& \quad \times \Biggl\{ \biggl[ \nu g^{\mu\rho} g^{\nu\sigma} (q_1-q_2)^\tau 
-  g^{\mu\rho} q_1^\nu  q_2^\sigma q_1^\tau  \\
&\quad + g^{\nu\rho}  \left(q_1^\mu  + q_2^\mu + \frac{Q_1^2}{\nu} q_2^\mu \right) q_2^\sigma q_1^\tau \biggr]  \frac{1}{M_{f_1}^2} F^{TT}_{f_1\gamma^*\gamma^*}(Q_1^2,0)  \\
&\quad -  g^{\nu\rho} \biggl(q_1^\mu +\frac{Q_1^2}{\nu} q_2^\mu \biggr) q_1^\sigma q_2^\tau  \frac{1}{M_{f_1}^2} F^{LT}_{f_1\gamma^*\gamma^*}(Q_1^2,0) \Biggr\}.
\end{aligned}
\end{equation}
To describe the fusion of one virtual photon and one real photon, only two independent TTFs $F^{TT,LT}_{f_1\gamma^*\gamma^*}(Q_1^2,0)$ are needed, while the third one, $F^{TL}_{f_1\gamma^*\gamma^*}(Q_1^2,0)$, decouples in the single virtual case~\cite{Pascalutsa:2012pr}. 
The two independent helicity amplitudes describing the  $\gamma^* \gamma \rightarrow f_1(1285)$ transition are then given by: 
\begin{equation}\label{Eq:HelAmpgagaF1}
\begin{aligned}
\mathcal{M}_{f_1 \gamma^* \gamma}^{\lambda_1=\lambda_2= 1, \Lambda_{f_1}=0} & = e^2 \frac{\nu Q_1^2}{M_{f_1}^3} F_{f_1 \gamma^* \gamma^*}^{T T}\left(Q_1^2,0\right), \\
\mathcal{M}_{f_1 \gamma^* \gamma}^{\lambda_1=0, \lambda_2=1, \Lambda_{f_1}=-1} & = -e^2 \frac{\nu Q_1}{M_{f_1}^2} F_{f_1 \gamma^* \gamma^*}^{LT}\left(Q_1^2,0\right). 
\end{aligned}
\end{equation}
One thus notices that the polarized cross section $\sigma_{TT}$ for the fusion of two transversely polarized photons is suppressed by a factor $Q_1^2/(2M_{f_1}^2)$ in comparison with the cross section $\sigma_{LT}$ in which the first photon is longitudinally polarized, in accordance with the Landau-Yang theorem~\cite{Landau:1948kw,Yang:1950rg}. This feature will be helpful to determine the TFF $F^{TL}_{f_1\gamma^*\gamma^*}(Q_1^2,0)$ nearly model-independently, as discussed in Ref.~\cite{Ren:2024nuf}.

Currently, the $f_1(1285)$ TFFs are not well determined due to the lack of the direct experimental data. 
The dipole form of $f_1(1285)$ TFFs, which is derived in the quark model~\cite{Schuler:1997yw}, has been widely used in e.g. Refs.~\cite{L3:2001cyf,Pauk:2014rta}. Melnikov and Vainshtein~\cite{Melnikov:2003xd} proposed a parametrization of the $f_1$ TFFs constrained by the large-$N_c$ and operator product expansion arguments, which was subsequently anti-symmetrized in Ref.~\cite{Jegerlehner:2017gek} to satisfy the Landau-Yang suppression. Recently, the $f_1(1285)$ TFFs have been studied using the resonance chiral theory~\cite{Roig:2019reh} and the holographic models~\cite{Leutgeb:2019gbz}. For high virtualities, the asymptotic behavior of $f_1$ TFFs has been derived from the light-cone expansion~\cite{Hoferichter:2020lap}. Incorporating this high-energy constraint, a vector meson dominance inspired parametrization of the $f_1(1285)$ TFFs have been proposed in Refs.~\cite{Zanke:2021wiq,Hoferichter:2023tgp} through a global fit of all relevant data. 

For the sake of providing cross section estimates in the present work, we employ the quark model relation to determine $F^{TT}_{f_1\gamma^*\gamma^*}(Q_1^2,0)$ and $F^{LT}_{f_1\gamma^*\gamma^*}(Q_1^2,0)$, 
\begin{equation}\label{Eq:f1TFFs}
	F^{TT}_{f_1\gamma^*\gamma^*}(Q_1^2,0)=- F^{LT}_{f_1\gamma^*\gamma^*}(Q_1^2,0) = -\frac{F^{LT}_{f_1\gamma^*\gamma^*}(0,0)}{(1+Q_1^2/\Lambda_{f_1}^2)^2},
\end{equation}
where we use the dipole form to parameterize the TFFs, which satisfies the asymptotic behavior $1/Q_1^4$ at large virtuality, as given in~\cite{Hoferichter:2020lap}. 
The dipole mass $\Lambda_{f_1}$, which can be adjusted according to the relevant data, is fixed as $\Lambda_{f_1}=1040$ MeV, estimated by the L3 Collaboration~\cite{L3:2001cyf}.  
As to the normalization $F^{LT}_{f_1\gamma^*\gamma^*}(0,0)$, it is conventional to define an equivalent two-photon decay width of $f_1(1285)$ as 
\begin{equation}
\begin{aligned}
\tilde{\Gamma}_{f_1 \rightarrow \gamma \gamma} & \equiv \lim _{Q_1^2 \rightarrow 0} \frac{M_{f_1}^2}{2 Q_1^2} \Gamma\left(f_1 \rightarrow  \gamma_L^* \gamma_T \right) \\
& =\frac{\pi \alpha^2}{4} M_{f_1} \frac{1}{3}\left[F_{f_1 \gamma^* \gamma^*}^{LT}(0,0)\right]^2.
\end{aligned}
\end{equation}
The equivalent $f_1(1285)\to \gamma \gamma$ decay width has been estimated by the L3 Collaboration as~\cite{L3:2001cyf}:
\begin{equation}
\tilde{\Gamma}_{f_1\to \gamma\gamma} = 3.5\pm 0.8~\mathrm{keV}, 
\end{equation}
resulting in the normalization 
\begin{equation}
	F_{f_1 \gamma^* \gamma^*}^{LT}(0,0) = 0.44 \pm 0.05. 
\end{equation}

\subsection{$\gamma^*\gamma \to a_0(980)^\pm \pi^\mp \to \pi^+\pi^-\eta$ channel}

To obtain the amplitude of the $\gamma^*\gamma \to f_1(1285) \to a_0(980)^\pm \pi^\mp \to \pi^+\pi^-\eta$ channel, we need two further vertices: $f_1\to a_0 \pi$ and $a_0\to \pi \eta$. The Lagrangian to describe the $f_1(1285)$ decay to $a_0(980) \pi$ is given by~\footnote{Note that the expression of $\mathcal{L}_{f_1a_0\pi} = g_{f_1 a_0 \pi} f_1^\mu \partial_\mu \bm{\pi} \cdot \bm{a}_0$ is equivalent to Eq.~\eqref{Eq:Lf1a0pi} if the on-shell condition applies.} 
\begin{equation}~\label{Eq:Lf1a0pi}
		\mathcal{L}_{f_1a_0\pi} = \frac{g_{f_1 a_0 \pi}}{2} f_1^\mu \bigl( \partial_\mu \bm{\pi} \cdot \bm{a}_0 -  \bm{\pi} \cdot \partial_\mu \bm{a}_0 \bigr), 
\end{equation}
where $\bm{\pi}$ and $\bm{a}_0$ denote the isovector $\pi$ and $a_0$ fields, and $f_1^\mu$ is the $f_1(1285)$ vector field. 
The dimensionless coupling $g_{f_1a_0\pi}$ is determined by the partial decay width of $f_1(1285)\to a_0(980) \pi$, which is a $p$-wave decay process,
\begin{equation}\label{Eq:f1TOa0pi}
\begin{aligned}
	\Gamma_{f_1\to a_0\pi} &= \Gamma_{f_1\to a_0^+\pi^-} + \Gamma_{f_1\to a_0^-\pi^+} + \Gamma_{f_1\to a_0^0\pi^0} \\
	&= \frac{g_{f_1 a_0\pi}^2}{8\pi M_{f_1}^2} \left[q_{f_1\to a_0 \pi}(M_{f_1}^2)\right]^3, 
\end{aligned}
\end{equation}
with the center-of-mass (cm) momentum in the rest frame of the $f_1(1285)\to a_0(980) \pi$ decay process, 
\begin{equation}
	q_{f_1\to a_0\pi}(W^2) = \frac{\lambda^{1/2}(W^2, M_{a_0}^2, m_\pi^2)}{2 W},
\end{equation}
and $\lambda$ being the K\"{a}ll\'{e}n triangle function $\lambda(x,y,z)\equiv x^2+y^2+z^2-2xy-2xz-2yz$.  
The partial decay width 
\begin{equation}
\Gamma^\mathrm{Exp.}_{f_1\to a_0\pi}\simeq 8.63\pm 1.00~\mathrm{MeV},
\end{equation}
is obtained using the branching ratio $\mathcal{B}(f_1(1285)\to a_0 \pi) = 38\pm 4\%$ averaged by PDG~\cite{ParticleDataGroup:2024cfk}. The coupling constant $g_{f_1 a_0 \pi}$ is then estimated via Eq.~\eqref{Eq:f1TOa0pi} as: 
\begin{equation}
	g_{f_1a_0\pi}\simeq 5.26\pm 0.69. 
\end{equation}

The effective $a_0\pi\eta$ interaction is described by the Lagrangian 
\begin{equation}
	\mathcal{L}_{a_0 \eta \pi} = g_{a_0\eta \pi} \bm{a}_0 \cdot \partial_\mu \bm{\pi} \, \partial^\mu \eta , 
\end{equation}
where we absorb the $m_\pi^2$ term~$\sim \bm{a_0}\cdot\bm{\pi}\,\eta$ in the chiral Lagrangian~\cite{Ecker:1988te} into the effective coupling $g_{a_0\eta \pi}$~\footnote{Note that in Ref.~\cite{Rudenko:2017bel} the structure $g_{a_0\eta\pi} \bm{a_0}\cdot\bm{\pi}\,\eta$ is used to describe the $a_0\pi\eta$ vertex, which however does not obey the chiral limit behavior.}.

The coupling $g_{a_0\eta \pi}$  
can be fixed via the decay width of the $a_0(980)\to \eta \pi$:
\begin{equation}
\begin{aligned}
	\Gamma_{a_0\to\eta\pi} &= \Gamma_{a_0^+\to\eta\pi^+}  = \Gamma_{a_0^-\to\eta\pi^-}  \simeq \Gamma_{a_0^0\to\eta\pi^0}  \\
	& =\frac{g_{a_0\eta\pi}^2 }{8\pi M_{a_0}^2} \, q_{a_0\to \eta \pi} (M_{a_0}^2) \\
	&\quad \times   \biggl\{M_{a_0}^2\, [q_{a_0\to \eta \pi} (M_{a_0}^2) ]^2 + m_\pi^2 m_\eta^2 \biggr\}.
\end{aligned}
\end{equation}
There are five decay modes of  the $a_0(980)$ resonance: $a_0(980)\to\eta\pi$,  $a_0(980)\to K\bar{K}$, $a_0(980)\to \eta'\pi$, $a_0(980)\to \gamma\gamma$,  and $a_0(980)\to e^+e^-$, as listed in the PDG review~\cite{ParticleDataGroup:2024cfk}. 
Assuming that the total $a_0(980)$ width comes from the first two dominant channels, the partial decay width of $a_0\to \eta \pi$ is given as  
\begin{equation}\label{Eq:a0etapiDecay}
	\Gamma_{a_0\to \eta\pi}^\mathrm{Exp.} \simeq \frac{1}{1.172}\Gamma_{a_0}^\mathrm{Exp.} = 85.32\%\, \Gamma_{a_0}^\mathrm{Exp.},
\end{equation}
where we have used the PDG-averaged branching ratio $\Gamma(a_0\to K\bar{K})/\Gamma(a_0\to \eta \pi) = 0.172\pm 0.019$~\cite{ParticleDataGroup:2024cfk}.  
We note that the position and width of the $a_0(980)$ resonance have a large model dependence. In this work, we employ the latest result for the $a_0(980)$ parameters, reported by the Belle Collaboration~\cite{Belle:2009xpa}: 
\begin{equation*}
\begin{aligned}
	M_{a_0}^\mathrm{Exp.} &= 982.3\left(^{+0.6}_{-0.7}\right)\left(^{+3.1}_{-4.7}\right) ~\mathrm{MeV},\\
	 \Gamma_{a_0}^\mathrm{Exp.} &= 75.6\left(^{+1.6}_{-1.6}\right) \left(^{+17.4}_{-10.0} \right)~\mathrm{MeV},
\end{aligned}
\end{equation*} 
where the first (second) bracket refers to the statistic (systematic) uncertainty. This then leads to the coupling constant:
\begin{equation}
g_{a_0\eta \pi} = 6.85\pm 0.65~\mathrm{GeV}^{-1}.
\end{equation}

\begin{table}[t]
\caption{Decay channels of  $f_1(1285)$~\cite{ParticleDataGroup:2024cfk}.}
\label{Tab:f1decay}
\renewcommand*{\arraystretch}{1.4}
\begin{tabular*}{\columnwidth}{@{\extracolsep{\fill}}lrr}
\hline \hline Branch ratio & & Value  \\
\hline 
$\mathcal{B}(f_1 \to a_0 \pi )$ &~~~~~ & $38 \pm 4\, \%$ \\
$\mathcal{B}(f_1 \to \pi^0\pi^0 \pi^+ \pi^-)$ & & $21.8 \pm 1.3\,\%$ \\
$\mathcal{B}(f_1 \to \eta \pi \pi )$ [excluding $a_0 \pi$] & & $14 \pm 4\, \%$ \\
$\mathcal{B}(f_1 \to \rho^0 \pi^+\pi^-)$ & & $10.9 \pm 0.6\,\%$ \\
$\mathcal{B}(f_1 \to K\bar{K}\pi )$ &  & $9.0 \pm 0.4\, \%$  \\
$\mathcal{B}(f_1 \to \gamma \rho^0 )$ & & $6.1 \pm 1.0\, \%$\\
\hline \hline
\end{tabular*}
\end{table}

Having specified all vertices, one can express the $s$-channel amplitude of Fig.~\ref{Fig:Feyn}(a) via the $a_0^\pm(980)\pi$ decay channel as
\begin{equation}
	\mathcal{M}_a = \mathcal{M}_{a_0^+\pi^-} + \mathcal{M}_{a_0^-\pi^+}  \, ,
\end{equation}
where 
\begin{widetext}
\begin{equation}
\begin{aligned}
	\mathcal{M}_{a_0^+\pi^-} &= i\,e^2\, g_{f_1a_0\pi}\, g_{a_0\eta\pi}  \, \varepsilon_\mu(q_1,\lambda_1) \varepsilon_\nu(q_2,\lambda_2)  \frac{\sum\varepsilon^{*\alpha}(P_{f_1},\Lambda_{f_1})\varepsilon^\beta(P_{f_1},\Lambda_{f_1})}{(q_1+q_2)^2-M_{f_1}^2+i M_{f_1}\Gamma_{f_1}(P_{f_1}^2)} \\
	&\times  \sqrt{\frac{D_1\bigl[q_{f_1 \rightarrow \gamma^* \gamma}(s) R_{f_1}\bigr]}{D_1\bigl[q_{f_1 \rightarrow \gamma^* \gamma}(M_{f_1}^2) R_{f_1}\bigr]}} \sqrt{\frac{D_1\bigl[q_{f_1 \rightarrow a_0 \pi} (s) R_{f_1}\bigr]}{D_1\bigl[q_{f_1 \rightarrow a_0 \pi} (M_{f_1}^2) R_{f_1}\bigr]}} \\
	&\times  \epsilon_{\rho\sigma\tau\alpha}   \, (p_{\pi^-})_\beta  \frac{p_{\pi^+}\cdot p_\eta}{p_{a_0^+}^2-M_{a_0}^2+i M_{a_0}\Gamma_{a_0}(p_{a_0^+}^2)} \Biggl\{ \biggl[ \nu g^{\mu\rho} g^{\nu\sigma} (q_1-q_2)^\tau 
-  g^{\mu\rho} q_1^\nu  q_2^\sigma q_1^\tau  \\
& + g^{\nu\rho}  \left(q_1^\mu  + q_2^\mu + \frac{Q_1^2}{\nu} q_2^\mu \right) q_2^\sigma q_1^\tau \biggr]  \frac{1}{M_{f_1}^2} F^{TT}_{f_1\gamma^*\gamma^*}(Q_1^2,0) -  g^{\nu\rho} \biggl(q_1^\mu +\frac{Q_1^2}{\nu} q_2^\mu \biggr) q_1^\sigma q_2^\tau  \frac{1}{M_{f_1}^2} F^{LT}_{f_1\gamma^*\gamma^*}(Q_1^2,0) \Biggr\},\\
	\mathcal{M}_{a_0^-\pi^+} &=\left.\mathcal{M}_{a_0^+\pi^-}\right|_{p_{\pi^+}\leftrightarrow p_{\pi^-}},
\end{aligned}
\end{equation}
with the $f_1(1285)$ momentum $P_{f_1}\equiv q_1+q_2$ and the $a_0(980)^+$ momentum $p_{a_0^+}\equiv p_{\eta} + p_{\pi^+}$. The Blatt-Weisskopf barrier factor~\cite{Blatt:1952ije} $D_1(x)=1/(x^2+1)$ is taken into account in the amplitude for the $p$-wave decay of $f_1(1285) \to a_0(980)^{+} \pi^-$. The constant $R_{f_1}$ is the effective range parameter for the $f_1(1285)$ resonance, which we fix as $R_{f_1}=3.0$ GeV$^{-1}$, consistent with the range reported in ~\cite{Belle:2009xpa}. 
The energy-dependent width of the intermediate $f_1(1285)$ resonance is parameterized as:
\begin{equation}\label{Eq:f1width}
\begin{aligned}
 \Gamma_{f_1}(P_{f_1}^2) 
=   \,\Gamma_{f_1}(M_{f_1}^2) \Biggl\{ & \mathcal{B}(f_1\to a_0\pi) \frac{M_{f_1}}{\sqrt{s}} \left[\frac{q_{f_1 \rightarrow a_0 \pi} (s)}{q_{f_1 \rightarrow a_0 \pi}(M_{f_1}^2)}\right]^3  \frac{D_1\left[q_{f_1 \rightarrow a_0 \pi} (P_{f_1}^2) R_{f_1}\right]}{D_1\left[q_{f_1 \rightarrow a_0 \pi} (M_{f_1}^2) R_{f_1}\right]} \Theta\bigl(P_{f_1}^2 - (M_{a_0}+m_{\pi})^2\bigr) \\
& +  \mathcal{B}(f_1\to \gamma \rho^0) \frac{M_{f_1}}{\sqrt{P_{f_1}^2}}  \frac{q_{f_1 \rightarrow \gamma\rho^0} (P_{f_1}^2)}{q_{f_1 \rightarrow \gamma\rho^0}(M_{f_1}^2)}   \Theta\bigl(P_{f_1}^2 - (M_{\rho^0})^2\bigr) \\
& + \biggl[\mathcal{B}(f_1\to \pi^0\pi^0\pi^+ \pi^-) + \mathcal{B}(f_1\to \eta\pi \pi) +  \mathcal{B}(f_1\to \rho^0\pi^+ \pi^-) +  \mathcal{B}(f_1\to K\bar{K}\pi ) \biggr] \frac{P_{f_1}^2}{M_{f_1}^2} \Biggr\},
\end{aligned}
\end{equation}
where the branching ratios of $f_1(1285)$ resonance are listed in Table~\ref{Tab:f1decay}. 
For the three- and four-body decay modes, we follow Ref.~\cite{Belle:2009xpa} to parameterize their contribution to the energy-dependent width as the linear term $P_{f_1}^2/M_{f_1}^2$.  

Correspondingly, the energy-dependent width of the $a_0(980)^+$ resonance is parameterized as: 
\begin{equation}
\begin{aligned}
  \Gamma_{a_0}(p_{a_0^+}^2)  = \Gamma_{a_0}(M_{a_0}^2) \Biggl\{ &\mathcal{B}(a_0\to \pi\eta) \frac{M_{a_0}}{\sqrt{p_{a_0^+}^2}} 
  \frac{q_{a_0\to \pi\eta}(p_{a_0^+}^2)}{q_{a_0\to \pi\eta}(M_{a_0}^2)}   \Theta\bigl(P_{a_0}^2-(m_\pi+m_\eta)^2\bigr) \\
  &+ \mathcal{B}(a_0\to K\bar{K}) \frac{M_{a_0}}{\sqrt{p_{a_0^+}^2}} 
  \frac{q_{a_0\to K\bar{K}}(p_{a_0^+}^2)}{q_{a_0 \to K\bar{K}}(M_{a_0}^2)}    \Theta\bigl(p_{a_0^+}^2-(2m_K)^2\bigr) \Biggr\},
\end{aligned}
\end{equation}
\end{widetext}
where the branching ratios $\mathcal{B}(a_0\to \pi\eta)$ and $\mathcal{B}(a_0\to K\bar{K})$ are $85.32\%$ and $14.68\%$, respectively, as determined by Eq.~\eqref{Eq:a0etapiDecay}. 

The rest-frame momenta of the considered channels are given by 
\begin{align}\label{Eq:momenta}
		q_{f_1\to\gamma^*\gamma}(s) & = \frac{\lambda^{1/2}(s, -Q_1^2, 0)}{2\sqrt{s}},\\ 
 	q_{f_1\to a_0 \pi}(s) &= \frac{\lambda^{1/2}(s, p_{a_0^+}^2,m_\pi^2)}{2\sqrt{s}}, \\
 	q_{a_0\to\pi\eta}(p_{a_0^+}^2) &= \frac{\lambda^{1/2}(p_{a_0^+}^2, m_\eta^2, m_\pi^2)}{2\sqrt{p_{a_0^+}^2}}.
\end{align}

\subsection{$\gamma^*\gamma\to \sigma/f_0(500) \eta \to \pi^+\pi^-\eta$ channel}

We next discuss the $f_1(1285) \to \sigma \eta$ channel. 
The effective Lagrangian to describe the $f_1\sigma \eta$ interaction vertex is given by
\begin{equation}
	\mathcal{L}_{f_1\sigma \eta} = \frac{g_{f_1\sigma\eta}}{2} f_1^\mu \bigl( \partial_\mu \eta\, \sigma - \eta\, \partial_\mu \sigma \bigr), 
\end{equation}
and the effective coupling constant $g_{f_1\sigma \eta}$ can be determined from the partial decay width of $f_1(1285)\to \sigma/f_0(500) \eta$. We use the following expression 
\begin{align}\label{eq:momenta2}
&\Gamma_{f_1\to \sigma\eta} = \frac{1}{\Delta M_\sigma}\int_{M_\sigma^\mathrm{min}}^{M_\sigma^\mathrm{max}} dM_\sigma \frac{g_{f_1 \sigma\eta}^2}{24\pi M_{f_1}^2} \,q^3_{f_1\to \sigma \eta}(M_{f_1}^2),\nonumber\\
&q_{f_1\to \sigma \eta}(P_{f_1}^2) = \frac{\lambda^{1/2}(P_{f_1}^2, M_\sigma^2,m_\eta^2)}{2\sqrt{P_{f_1}^2}}\,,
\end{align}
by averaging over the mass range $M_{\sigma}$ from $M_\sigma^\mathrm{min}=400$ MeV to $M_{\sigma}^\mathrm{max}=550$ MeV, with $\Delta M_\sigma=150$ MeV. 
According to the PDG review~\cite{ParticleDataGroup:2024cfk}, the branching ratio of $f_1(1285)\to \eta \pi \pi$ excluding the $f_1\to a_0\pi$ contribution is averaged as $14\pm 4\%$. 
We use this result as the maximum value of $\Gamma_{f_1\sigma\eta}$ leading to the coupling constant 
\begin{equation}
g_{f_1\to \sigma\eta}\simeq 2.62 \pm 0.39. 
\end{equation}
In this way, the effects of the higher resonances, such as $f_0(980)$, $f_2(1270)$, in the same production mechanism as the $\sigma\eta$ channel are considered effectively.   
In order to estimate the contribution from the $\gamma^*\gamma \to \sigma \eta \to \pi^+\pi^-\eta$ process, we first describe the amplitude for the $\gamma^*\gamma \to \sigma \eta$ subprocess as: 
\begin{widetext} 
\begin{equation}\label{Eq:sigmaeta}
\begin{aligned}
\mathcal{M}_{\sigma\eta} & = i\, e^2 g_{f_1\sigma\eta}\, \varepsilon_{\mu}(q_1,\lambda_1) \,\varepsilon_{\nu}(q_2,\lambda_2)\,\epsilon_{\rho\sigma\tau\alpha}\,(p_\eta)_\beta  \frac{\sum\varepsilon^{*\alpha}(P_{f_1},\Lambda_{f_1})\varepsilon^\beta(P_{f_1},\Lambda_{f_1})}{(q_1+q_2)^2-M_{f_1}^2+i M_{f_1}\Gamma_{f_1}(P_{f_1}^2)} \\
	& \times \sqrt{\frac{D_1\left[q_{f_1 \rightarrow \gamma^* \gamma}(P_{f_1}^2) R_{f_1}\right]}{D_1\left[q_{f_1 \rightarrow \gamma^* \gamma}(M_{f_1}^2) R_{f_1}\right]}} \sqrt{\frac{D_1\left[q_{f_1 \rightarrow \sigma \eta} (P_{f_1}^2) R_{f_1}\right]}{D_1\left[q_{f_1 \rightarrow \sigma \eta} (M_{f_1}^2) R_{f_1}\right]}}  \Biggl\{ \biggl[ \nu g^{\mu\rho} g^{\nu\sigma} (q_1-q_2)^\tau 
-  g^{\mu\rho} q_1^\nu  q_2^\sigma q_1^\tau  \\
&\quad + g^{\nu\rho}  \left(q_1^\mu  + q_2^\mu + \frac{Q_1^2}{\nu} q_2^\mu \right) q_2^\sigma q_1^\tau \biggr]  \frac{1}{M_{f_1}^2} F^{TT}_{f_1\gamma^*\gamma^*}(Q_1^2,0)  -  g^{\nu\rho} \biggl(q_1^\mu +\frac{Q_1^2}{\nu} q_2^\mu \biggr) q_1^\sigma q_2^\tau  \frac{1}{M_{f_1}^2} F^{LT}_{f_1\gamma^*\gamma^*}(Q_1^2,0) \Biggr\},
\end{aligned}
\end{equation} 
\end{widetext}
where we take the same form of the energy-dependent width of $f_1(1285)$ as given in Eq.~\eqref{Eq:f1width}. 
The Blatt-Weisskopf barrier factor $D_1(x)$ from the $p$-wave decays $f_1(1285)\to\gamma^*\gamma$ and $f_1(1285) \to \sigma\eta$ is taken into account with the momenta $q_{f_1\to \gamma^*\gamma}$ and $q_{f_1\to \sigma \eta}$ defined in Eqs.~\eqref{Eq:momenta} and ~(\ref{eq:momenta2}), respectively.

To describe the subsequent decay of $\sigma/f_0(500) \to \pi^+\pi^-$, we use the $s$-wave isospin $I=0$ Omn\`es function, 
\begin{equation}
\Omega(x)=\exp \left\{\frac{x}{\pi} \int_{4 m_\pi^2}^{\infty} \frac{d x^{\prime}}{x^{\prime}} \frac{\delta\left(x^{\prime}\right)}{x^{\prime}-x}\right\},
\end{equation}
 which accounts for the rescattering effects through the $\sigma/f_0(500)$ resonance. Since the Omn\`es function is normalized as $\Omega(0)=1$, one needs to introduce a coupling $C$, which has the dimension GeV$^{-1}$, to obtain the amplitude of the $\gamma^*\gamma \to \sigma \eta \to \pi^+\pi^-\eta$ process, 
 \begin{equation}
 	\mathcal{M}_b= C \, \mathcal{M}_{\sigma\eta}
 	\,\Omega(M_{\pi^+\pi^-}^2),
  \label{eq:C}
 \end{equation}
 where, in principle, the coupling $C$ is a function of $M_{\pi^+\pi^-}$. Due to the lack of experimental data of this process, here we take it as a constant for simplicity. Note that one needs to replace $M_\sigma$ appeared in Eq.~\eqref{Eq:sigmaeta} of the $\gamma^*\gamma\to\sigma\eta$ amplitude as the invariant mass $M_{\pi^+\pi^-}$ in  $\mathcal{M}_b$.
 
\begin{figure*}[t]
	\includegraphics[width=0.95\textwidth]{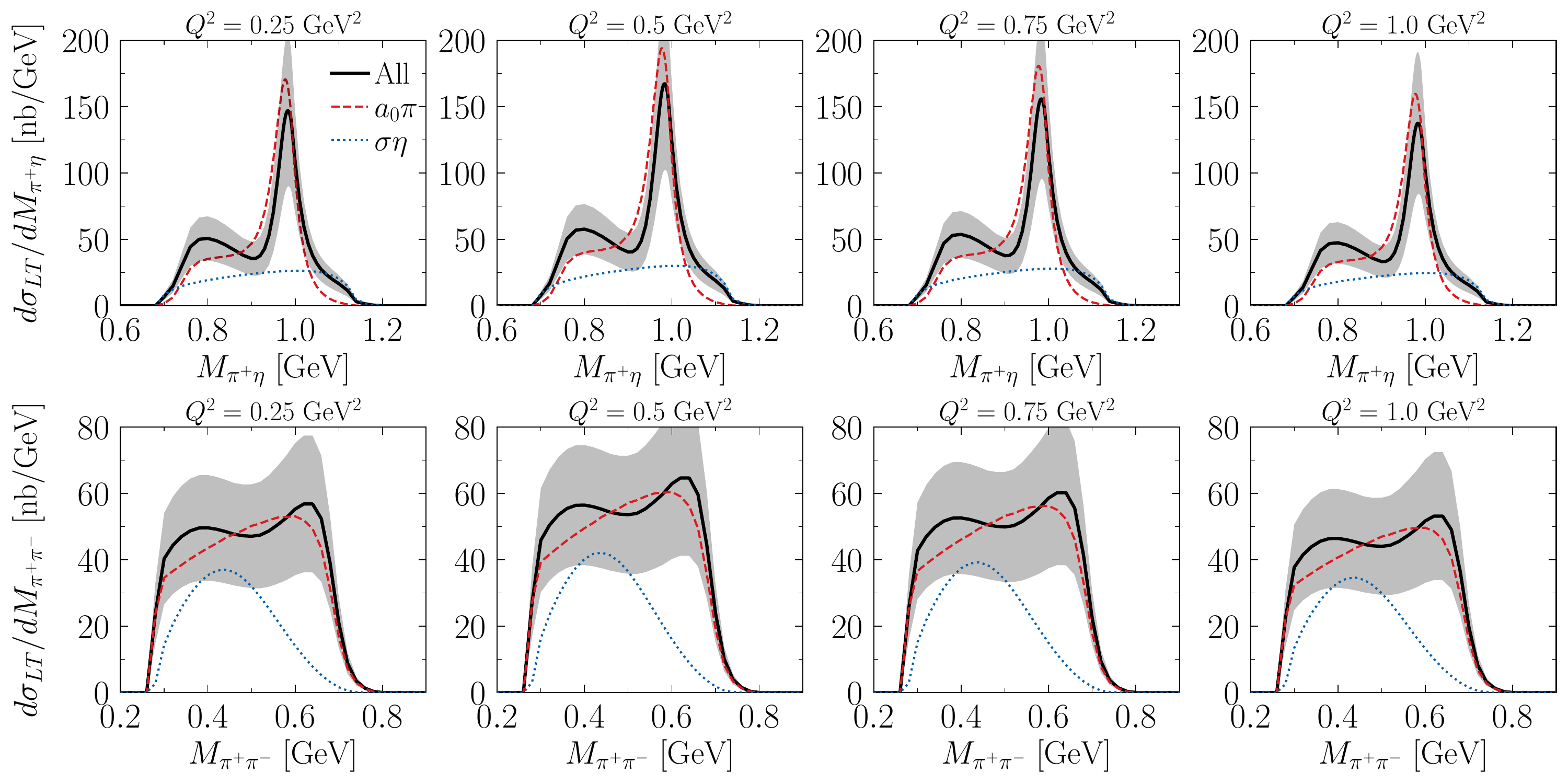}
	\caption{The predicted mass distributions $M_{\pi^+\eta}$ and $M_{\pi^+\pi^-}$ of the longitudinal-transverse cross section $\sigma_{LT}$ of the $\gamma^*\gamma \to \eta\pi^+\pi^-$ process for $Q^2=0.25$, $0.5$,  $0.75$, $1.0$ GeV$^2$. The black solid lines denote the total results of our model. The contributions from the $a_0^\pm(980) \pi^\mp$ and $\sigma\eta$ channels are indicated by the red dashed and blue dotted curves. The gray bands present the uncertainties of our results at $1\sigma$ level.}
	\label{Fig:LTinvmass}
\end{figure*}

 To determine the coupling $C$, we perform a matching at the level of the polarized total cross sections between the $\gamma^*\gamma\to \sigma \eta$ and $\gamma^*\gamma\to\pi^+\pi^- \eta$ processes.  By taking into account the $M_\sigma$ uncertainty in the $\gamma^*\gamma \to \sigma \eta$ process, we have the following relation by including the branching ratio $\mathrm{Br}(\sigma\to\pi^+\pi^-)=2/3$,
 \begin{equation}
 \begin{aligned}
 \sigma_{++}^{\gamma^*\gamma\to \eta \pi^+\pi^-} &=
   \frac{\mathrm{Br}(\sigma \to\pi^+\pi^-) }{\Delta M_\sigma} \int_{M_\sigma^\mathrm{min}}^{M_\sigma^\mathrm{max}} d M_\sigma 
   \sigma_{++}^{\gamma^*\gamma\to \sigma \eta}(M_\sigma), \\
  \sigma_{0+}^{\gamma^*\gamma\to \eta \pi^+\pi^-} &=
   \frac{\mathrm{Br}(\sigma \to\pi^+\pi^-) }{\Delta M_\sigma} \int_{M_\sigma^\mathrm{min}}^{M_\sigma^\mathrm{max}} d M_\sigma 
   \sigma_{0+}^{\gamma^*\gamma\to \sigma \eta}(M_\sigma),   
\end{aligned}
\end{equation}
where we average the two-body helicity cross sections by integrating over the $\sigma$ mass from $400$ MeV to $550$ MeV. 
 Since $C$ is approximated as a constant, we determine for $s=M_{f_1}^2$ and the virtuality $Q_1^2=0.5$ GeV$^2$, which is the virtuality corresponding with the  BESIII measurements discussed below, constraining its absolute value as $|C| \simeq 13.88$ GeV$^{-1}$.

 \begin{figure*}[t]
	\includegraphics[width=0.95\textwidth]{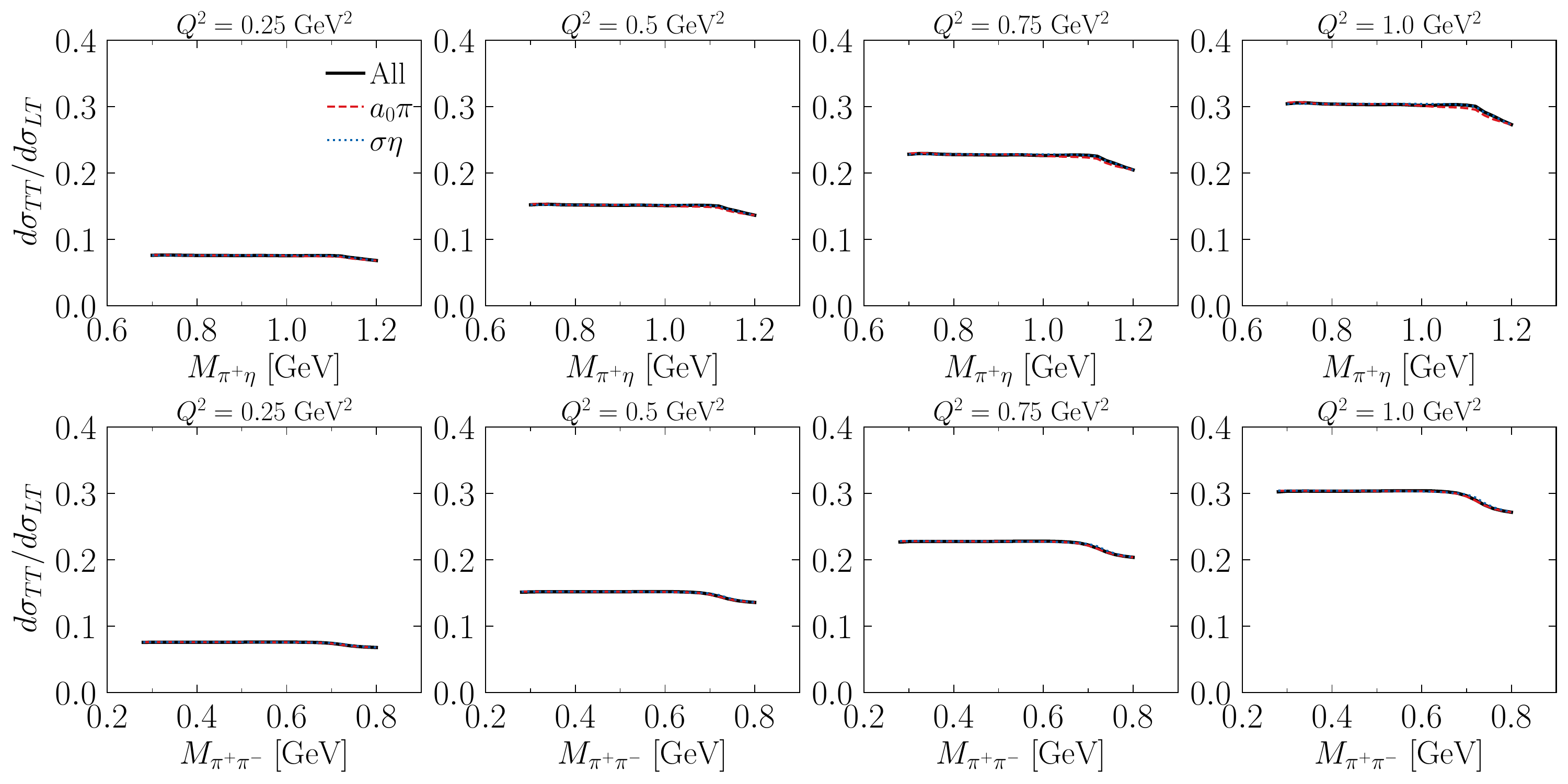}
	\caption{The ratio $d\sigma_{TT}/d\sigma_{LT}$ of mass distributions $M_{\pi^+\eta}$ and $M_{\pi^+\pi^-}$ of the  $\gamma^*\gamma \to \eta\pi^+\pi^-$ process for  $Q^2=0.25$, $0.5$,  $0.75$, $1.0$ GeV$^2$. The curve conventions are the same as in Fig.~\ref{Fig:LTinvmass}.} 
	\label{Fig:TToverLTinvmass}
\end{figure*}

 \section{Results and discussion}\label{SecIII}
In our model, tailored to describe the $f_1(1285)$ resonance excitation in the $\gamma^* \gamma \to \eta\pi^+\pi^-$ process, the amplitude can be written as sum of two subprocesses shown in Fig.~\ref{Fig:Feyn}
\begin{equation}
	\mathcal{M}^{\gamma^*\gamma \to \eta\pi^+\pi^-} = \mathcal{M}_a + \mathcal{M}_b\, ,
\end{equation}
where $\mathcal{M}_a$  and $\mathcal{M}_b$ correspond to the amplitudes of the $a_0^\pm \pi^\mp$ and $\sigma\eta$ channels, respectively. The relative phase between both amplitudes is not constrained by the above arguments, and will be chosen from the best fit to the available data as discussed further on. By introducing the helicity amplitudes $\mathcal{M}_{\lambda_1\lambda_2}$ of the $\gamma^*\gamma \to \eta\pi^+\pi^-$ process, one can evaluate the differential polarized cross sections of $\gamma^*\gamma\to \eta\pi^+\pi^-$
\begin{equation}\label{Eq:diffcrossX}
\begin{aligned}
	&\frac{d\sigma_{TT}}{d M_{\pi^+\eta}^2 dM_{\pi^+\pi^-}^2 d \cos\theta_{\pi^+\eta} d\phi} = \frac{\bigl( \mathcal{M}_{++}|^2+|\mathcal{M}_{-+}|^2 \bigr)}{128(2\pi)^4 s  \lambda^{1/2}(s,-Q_1^2,0)},  \\
&\frac{d\sigma_{LT}}{d M_{\pi^+\eta}^2 dM_{\pi^+\pi^-}^2 d\cos\theta_{\pi^+\eta}  d\phi} = \frac{ |\mathcal{M}_{0+}|^2}{64(2\pi)^4 s  \lambda^{1/2}(s,-Q_1^2,0)}, \\
\end{aligned}
\end{equation} 
with the angle $\theta_{\pi^+\eta}$ between $\bm{q}_1$ and the $\pi^+\eta$ momentum $\bm{P}_{\pi^+\eta}= \bm{p}_{\pi^+} + \bm{p}_{\eta}$ in the $\gamma^*\gamma$  c.m. frame. $\phi$ is defined  as the angle between the $\gamma^\ast \gamma \to a_0^+ \pi^-$  scattering plane and the $a_0^+(980)\to \pi^+\eta$ decay plane. 

Eqs.~\eqref{Eq:diffcrossX} allow to calculate the invariant mass distributions $d\sigma/M_{\pi^+\eta}$ and $d\sigma/M_{\pi^+\pi^-}$ as well as the angular distribution $d\sigma/d\cos\theta_{\pi^+\eta}$, which are functions of the total energy $W$ and the virtuality $Q_1^2$. 
Since our purpose in the present work is to provide the input for a Monte Carlo generator for the ongoing BESIII analysis of the $\gamma^*\gamma \to \eta\pi^+\pi^-$ process in the $f_1(1285)$ energy region, the invariant mass distributions and angular distributions are presented by averaging over the total energy bin   $1.22\leq W\leq 1.36$ GeV:
\begin{equation}
	\frac{d\sigma(Q_1^2)}{d M_{\pi^+\eta}} = \frac{1}{\Delta W} \int_{W_\mathrm{min}}^{W_\mathrm{max}} dW  \frac{d\sigma (W,Q_1^2)}{d M_{\pi^+\eta}}, 
\end{equation}
with $W_\mathrm{min}=1.22$ GeV and $W_\mathrm{max}=1.36$ GeV. An analogous averaging procedure applies to the calculations of $d\sigma/dM_{\pi^+\pi^-}$ and  $d\sigma/d\cos\theta_{\pi^+\eta}$.

We first present the invariant mass distributions using the parameterization for the $f_1(1285)$ TFFs given by Eq.~\eqref{Eq:f1TFFs}. We note that the lineshapes of the transverse-transverse ($TT$) and the longitudinal-transverse ($LT$) cross sections exhibit very similar behavior, apart from their magnitude. We therefore firstly discuss the result of the $LT$ mass distributions in Fig.~\ref{Fig:LTinvmass} for photon virtualities $Q_1^2=0.25$, $0.5$, $0.75$, $1.0$~GeV$^2$ covering the range of forthcoming BESIII data for the $\gamma^*\gamma\to \eta\pi^+\pi^-$ reaction.  We show our results for destructive interference between the $a_0^\pm \pi^\mp$ and $\sigma\eta$ amplitudes, resulting in a negative value for the coupling $C$ in Eq.~(\ref{eq:C}), i.e. $C \simeq -13.88$ GeV$^{-1}$. 
 In Fig.~\ref{Fig:LTinvmass} and the following, we also provide the error bands of our predictions, which result from error propagation of the couplings given by Eqs.~(11,16,20,30), which typically are known with around $10\%$ accuracy.
The resulting $M_{\pi^+\eta}$ mass distribution is consistent with the feature of L3 data: a clear $a_0^+(980)$ peak and a small shoulder on the lower side of the  $a_0^\pm(980)$ resonance peak, appearing as a kinematic reflection,  as shown in Fig.~(4)-c of Ref.~\cite{L3:2001cyf}. 
We note that this shoulder is further enhanced as a small bump through the interference with the $\sigma\eta$ channel. 
However the information provided by the L3 data is rather limited, as the $M_{\pi^+\eta}$ mass spectrum is only presented within the large bin of $Q^2$: $0.1 \sim 6.0$ GeV$^2$. Furthermore, there is no information about the $M_{\pi^+\pi^-}$ mass distribution, which prevent us from performing a detailed comparison with the L3 data.

\begin{figure*}[t]
	\includegraphics[width=0.95\textwidth]{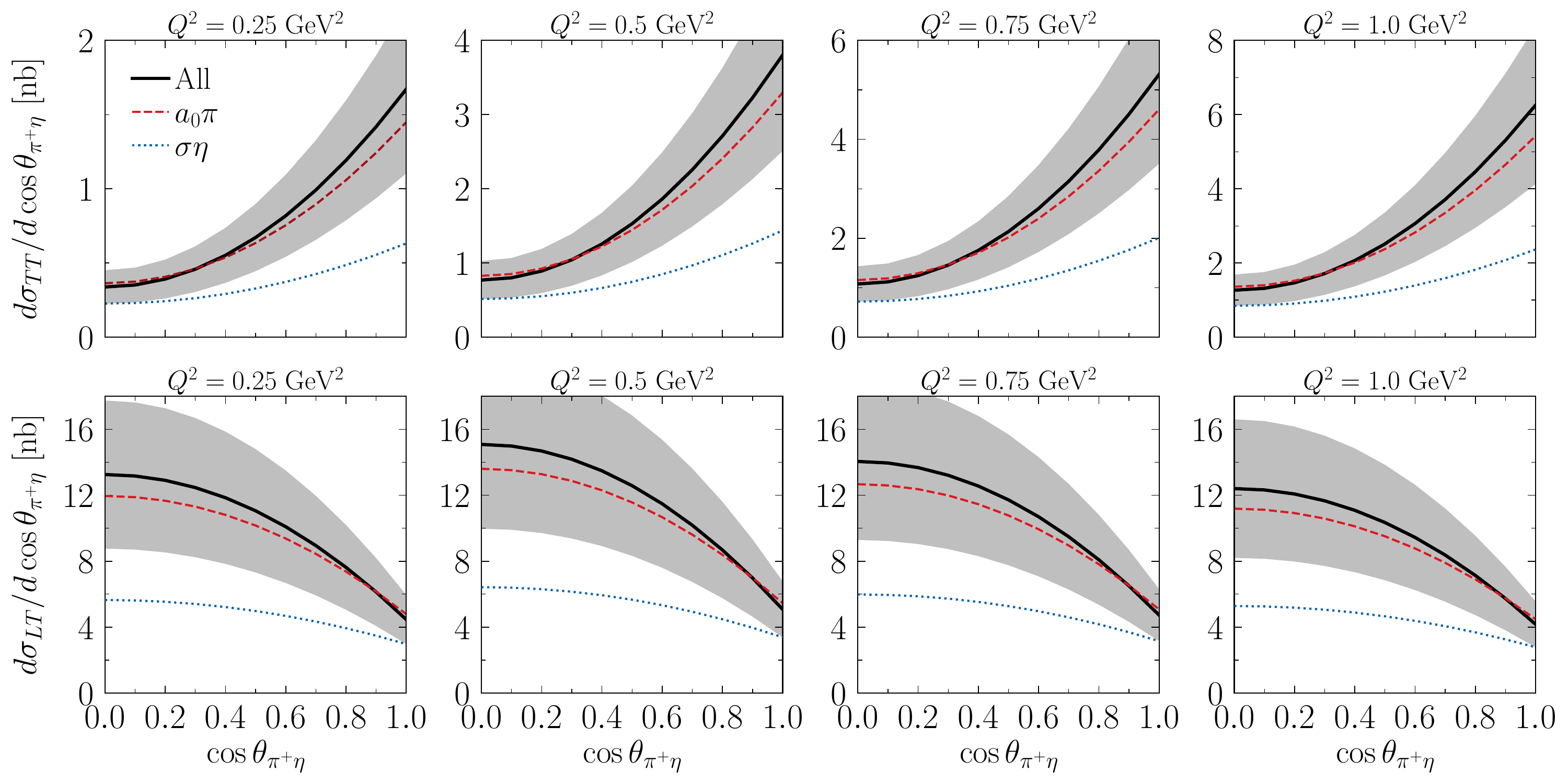}
	\caption{Predictions of the angular distribution for both the TT and the LT cross sections of the $\gamma^*\gamma \to \eta\pi^+\pi^-$ process for  $Q^2=0.25$, $0.5$,  $0.75$, $1.0$ GeV$^2$. The curve conventions are the same as Fig.~\ref{Fig:LTinvmass}.}
	\label{Fig:angledis}
\end{figure*}

For the $M_{\pi^+\pi^-}$ distribution, which is also shown in Fig.~\ref{Fig:LTinvmass}, the typical feature is the double peak structure around the edges of  the $M_{\pi^+\pi^-}$ physical region, which is produced by the destructive interference of both amplitudes. This fact indicates that the interference among the amplitudes of different channels is important for achieving the correct interpretation of experimental data. If there is no interference, or if the interference were constructive, only one peak around $M_{\pi^+\pi^-}=500$ MeV would appear in the $M_{\pi^+\pi^-}$ spectrum. The destructive interference between the $a_0^\pm \pi^\mp$ and $\sigma\eta$ amplitudes resulting in both  enhancements near the edges of the $M_{\pi^+\pi^-}$ mass distribution provides a definite prediction of the underlying reaction mechanism which can be tested with the forthcoming high statistics BESIII data.     

The $TT$ cross section is suppressed in magnitude relative to the $LT$ cross section over most of the $Q^2$ range. We show in Fig.~\ref{Fig:TToverLTinvmass} the $TT/LT$ cross section ratio for the same mass distributions as shown in Fig.~\ref{Fig:LTinvmass}. 
At $Q^2=0.25$ GeV$^2$, the $\sigma_{TT}$ result is around $10$ times smaller than the $\sigma_{LT}$ result. As $Q^2$ increases, this difference becomes smaller. When $Q^2$ reaches around $1.0$ GeV$^2$, the $TT/LT$ cross section ratio rises to about $0.3$. Such evolution is consistent with Eq.~\eqref{Eq:HelAmpgagaF1}, which results in 
$\sigma_{TT}$ suppressed by the factor $Q_1^2/(2M_{f_1}^2)$ in comparison with $\sigma_{LT}$.

To better illustrate the above mentioned effect of the destructive interference between both contributing channels, we also present the angular distributions $d\sigma_{TT}/d\cos\theta_{\pi^+\eta}$ and  $d\sigma_{LT}/d\cos\theta_{\pi^+\eta}$ in Fig.~\ref{Fig:angledis}. One can see that the $TT$ result, which is dominated by the $a_0^\pm(980)\pi^\mp$ channel, increases with increasing $\cos\theta_{\pi^+\eta}$. In contrast, the behavior of the $LT$ case is opposite, as $d\sigma_{LT}/d\cos\theta_{\pi^+\eta}$ decreases with increasing $\cos\theta_{\pi^+\eta}$. The order of magnitude difference between $TT$ and $LT$ is also seen again. 

\begin{figure*}[t]
	\includegraphics[width=0.95\textwidth]{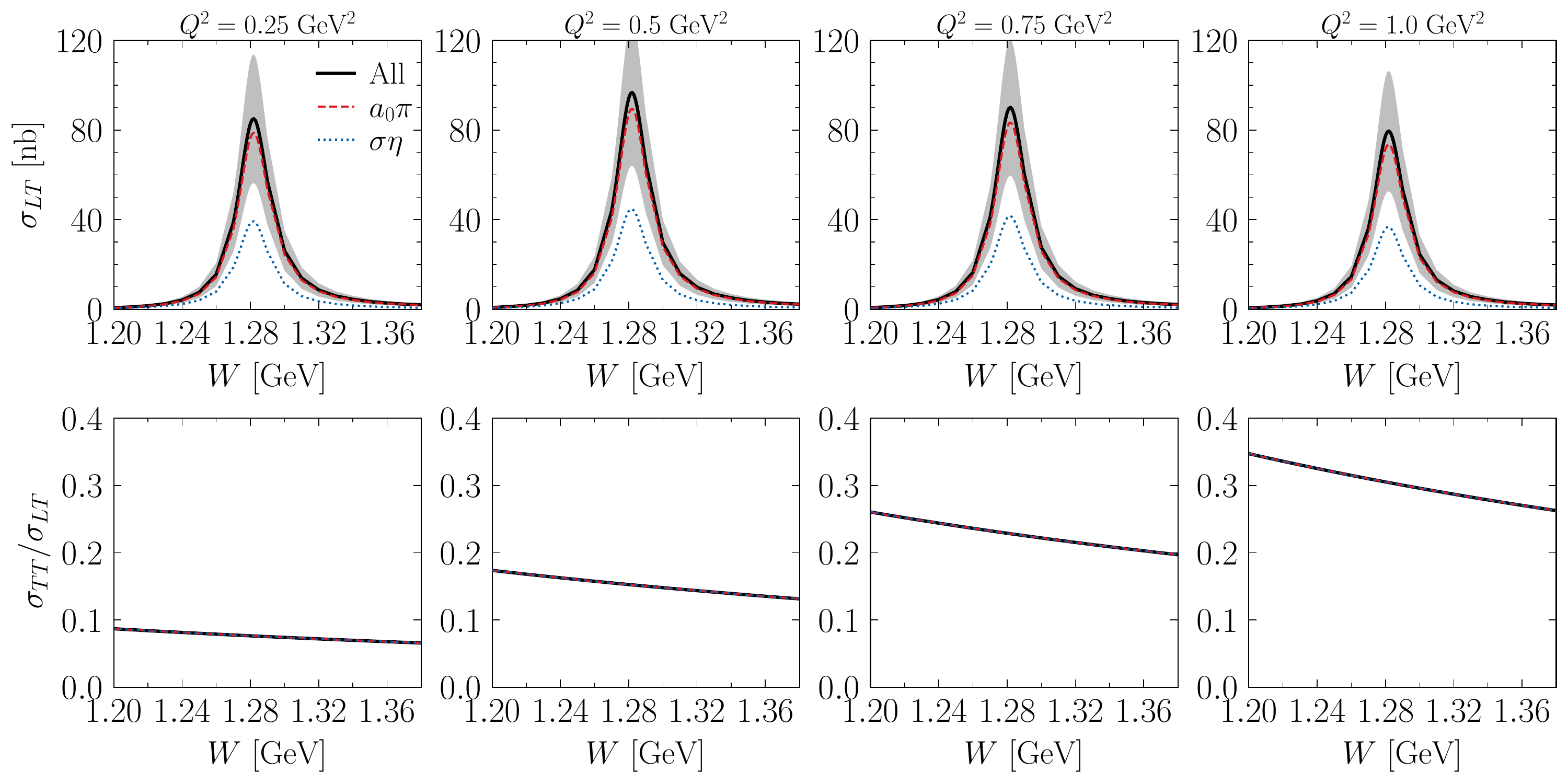}
	\caption{Predictions for total cross section $\sigma_{LT}$ (upper panel) and the ratio $\sigma_{TT}/\sigma_{LT}$ (lower panel) of the $\gamma^*\gamma\to\eta\pi^+\pi^-$ reaction with $Q^2=0.25$, $0.5$, $0.75$, and $1.0$ GeV$^2$. The curve notations are the same as in Fig.~\ref{Fig:LTinvmass}.}
	\label{Fig:TotXsec}
\end{figure*}

\begin{figure*}[t]
	\includegraphics[width=0.66\textwidth]{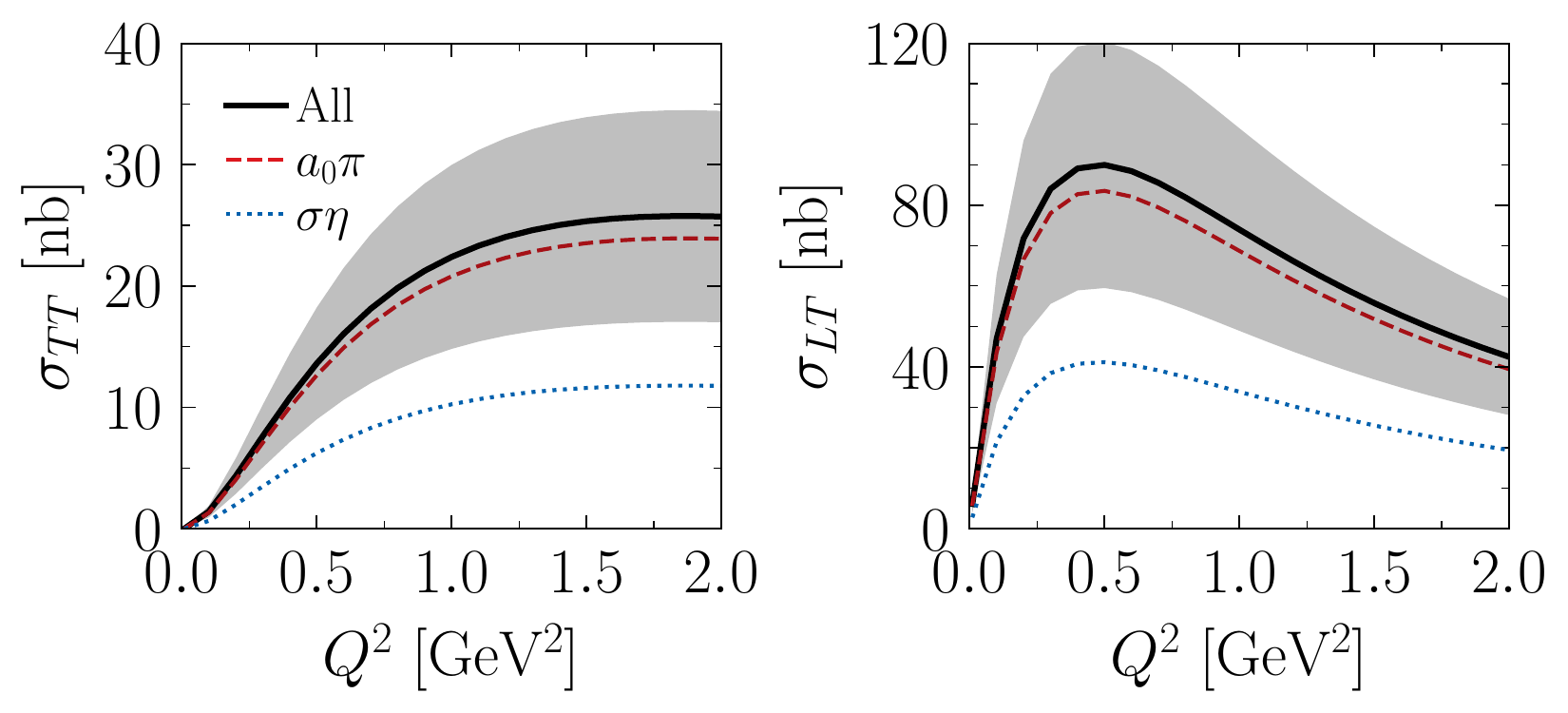}
	\caption{Predictions for the $Q^2$ dependence of the $\sigma_{TT,LT}$ total cross sections of the $\gamma^*\gamma\to\eta\pi^+\pi^-$ reaction for total energy $W=1.285$ GeV. The curve notations are the same as in Fig.~\ref{Fig:LTinvmass}.}
	\label{Fig:TotXsecQ2}
\end{figure*}

We present the $W$-dependence of the total cross section $\sigma_{LT}$ for the $\gamma^*\gamma\to \eta\pi^+\pi^-$ process in Fig.~\ref{Fig:TotXsec}. An interesting feature is that the contribution from the $a_0^\pm(980)\pi^\mp$ channel nearly saturates the entire cross section. This can be understood by the results of the invariant mass distributions and angular distributions as shown in Figs.~\ref{Fig:LTinvmass}-\ref{Fig:angledis}. However, the contribution from the $\sigma\eta$ channel cannot be neglected, because the destructive interference between the $a_0^\pm \pi^\mp$ and $\sigma \eta$ amplitudes is crucial for understanding the typical features of invariant mass distributions. Note that the effect of the BW factor was checked to be negligible in the cross sections and lies well within our shown error bands.

Finally, in Fig.~\ref{Fig:TotXsecQ2} we compare the $Q^2$ dependence of $\sigma_{TT}$ and $\sigma_{LT}$ total cross sections at the resonance position, i.e. for $W=1.285$ GeV. We note that the cross section $\sigma_{TT}$  increases gradually with increasing $Q^2$, whereas $\sigma_{LT}$ shows a different functional dependence: it reaches its maximum at $Q^2=0.5$ GeV$^2$ and gradually decreases as $Q^2$ increases. Such $Q^2$ dependence will be a sensitive observable in determining the $f_1(1285)$ TFFs from forthcoming high statistics BESIII data. 

\section{Summary and outlook}~\label{SecIV}
In conjunction with the current progress of the BESIII measurement on the diphoton fusion to the three-meson processes,  we have proposed a phenomenological model of the $\gamma^*\gamma\to \eta\pi^+\pi^-$ reaction tailored for the $f_1(1285)$ energy region. The channels from the $a_0(980)^{\pm}\pi^\mp$ and $\sigma/f_0(500)\eta$ intermediate states are considered within the effective Lagrangian approach. The destructive interference between both channels captures the main features of the invariant mass distributions of $\gamma^*\gamma\to \eta\pi^+\pi^-$. This finding indicates that the interference among the helicity amplitudes is essential in interpreting the experimental data. We also predict the angular distributions and polarized total cross sections of $\gamma^*\gamma\to \eta\pi^+\pi^-$ in the low $Q^2$ region of the BESIII measurement. 

Our work will serve as input to develop a Monte Carlo generator to assist the BESIII data analysis of the single tagged $e^+e^-\to e^+e^-\gamma^*\gamma\to e^+e^- \eta\pi^+\pi^-$ reaction. In this procedure, other possible mechanisms of $\gamma^*\gamma\to \eta\pi^+\pi^-$ may be included in order to achieve a good description of the expected high-statistics BESIII data. In the longer term, we aim to generalize this phenomenological study in a model-independent manner within a  dispersion framework once the relevant data become available. 

\acknowledgments 
We thank the useful discussions with Jan Muskalla and Dr. Christoph Florian Redmer about the current experimental analysis of the $\gamma^*\gamma\to \pi^+\pi^-\eta$ process.
This work was supported by the Deutsche Forschungsgemeinschaft (DFG, German Research Foundation), in part through the Research Unit (Photonphoton interactions in the Standard Model and beyond, Projektnummer 458854507—FOR 5327), and in part through the Cluster of Excellence (Precision Physics, Fundamental Interactions, and Structure of Matter) (PRISMA$^+$ EXC 2118/1) within the German Excellence Strategy (Project ID 39083149).

\bibliographystyle{apsrev4-2}
\bibliography{ref_gagatopipieta}

\end{document}